\documentclass{aa}
\usepackage{graphics}
\begin{document}
\thesaurus{03(09.03.2; 09.19.2; 13.18.1; 13.18.3}
\title{On the radio spectral index of galaxies} 

\author{Lisenfeld U.\inst{1} \and V\"olk H.J.\inst{2}}
\offprints{U. Lisenfeld}
\institute{IRAM, Avenida Divina Pastora 7, N.C., 18010 Granada,
Spain, e-mail: ute@iram.es \and
Max-Planck-Institut f\"ur Kernphysik, Saupfercheckweg 1,
D-69117 Heidelberg, Germany, e-mail: vlk@aposto.mpi-hd.mpg.de}
\date{Received 15 May 1999 / Accepted 3 December 1999}
\maketitle

\begin{abstract}

The radio emission of a galaxy consists of 
thermal bremsstrahlung, synchrotron emission  from discrete supernova remnants,
and diffuse synchrotron emission from cosmic ray electrons 
spread over the galactic disk and halo. Each of these components
has a different spectral index so that
the total radio spectral index of a galaxy depends sensitively on their 
relative contribution and on the processes shaping
the diffuse synchrotron emission. 
In the present paper we calculate the contribution of supernova remnants
to the total synchrotron emission and conclude that it is about
10\%. This moderate contribution has a noticeable effect on the
nonthermal spectral index, lowering it by $\simeq 0.1$ for steep
spectra. We then calculate the diffuse synchrotron emission in
two simple models, a diffusion and a convection model.
We find that the spatially integrated
nonthermal spectral index
is in general a poor diagnostic for the type of propagation or
the importance of energy losses. 
Spatially resolved radio data for the halo of galaxies are 
necessary in order to draw firm conclusions.
The steepening of the
spectrum away from the disk is a clear indication that 
synchrotron and inverse Compton losses are taking place 
during the propagation of cosmic ray electrons in the halo. 
In those galaxies for which spatially resolved data  for the
halos exist such a steepening has
been found. We conclude therefore that energy losses are 
generally important and that cosmic ray electrons
cannot freely escape from galaxies.

\end{abstract}

\section{Introduction}

The total radio emission of a galaxy consists of different 
components: The thermal radio emission, $P_\mathrm{therm}$, 
representing bremsstrahlung
from $\ion{H}{ii}$ regions and the nonthermal radio emission
which is synchrotron emission from cosmic ray  electrons (CREs).
As far as the nonthermal radio  emission, $P_\mathrm{nth}$, is concerned, 
a further division can be made by  
distinguishing between the diffuse radio emission, $P_\mathrm{diff}$, 
from CREs that are spread over the galactic disks and halos
and the  emission from discrete supernova remnants (SNRs), $P_\mathrm{SNR}$.
The three components possess very 
different spectral indices: Whereas the thermal
radio emission has a flat spectral index ($\propto \nu^{-0.1}$), the
nonthermal radio spectrum is steeper. The spectral index
of SNRs is on average $\approx \nu^{-0.5}$ (Green \cite{green})
and the diffuse radio emission can have, 
depending on the dominant process for the CR propagation, 
spectral indices\footnote
{Here and in the following we define the spectral index
$\alpha$ as $S(\nu)\propto \nu^{-\alpha}$}
, $\alpha$,
between about $0.5$ and $1.1$ which can, additionally, vary over the frequency
range. Therefore, the total radio spectral index 
depends very sensitively on the relative contribution of the 
different components.

The spectrum of the diffuse synchrotron emission 
is shaped by the physical processes that are 
characterizing the CR propagation, in particular the type of propagation
(diffusion or convection), of energy losses
(synchrotron, inverse Compton, adiabatic losses or bremsstrahlung), 
and the confinement of CREs. Here, we consider CREs  confined
to a galaxy if they loose their energy through combined
synchrotron and inverse Compton losses, down below the 
energy level corresponding to the observing frequency, before they
are able to leave the galactic halo. 
The question whether CREs are confined to galaxies or can
escape from the halos without suffering considerable energy losses
is important for the interpretation of the far-infrared (FIR)/radio
correlation. Different models have been proposed, explaining this
correlation by 
advocating either the situation that CREs are confined (V\"olk \cite{voelk_b};
Lisenfeld et al. \cite{lisenfeld_a})
or that they can escape more or less freely  (Chi \& Wolfendale \cite{chi}; 
Helou \& Bicay \cite{helou}; Niklas \& Beck \cite{niklas_a}).

The interpretation of the radio spectrum
in terms of the above mentioned processes is difficult because
they are hard to disentangle 
without spatially resolved observations. 
Furthermore, it is  necessary to
separate the diffuse radio emission from 
the thermal radio emission and the 
radio emission from SNRs.
The observational separation of the  three contributions 
based on their different spectral indices is difficult, especially for 
the radio emission of SNRs which has a  similar spectral
index as the diffuse synchrotron emission.
The separation of the thermal radio emission
is possible, if high-frequency radio data  
($\nu \approx  10-20$ GHz) are available, because its  
spectral index is very different. 

In this paper we are going to examine the influence 
of these various processes on the radio spectral index.
For this, we will start by
estimating the contribution of SNRs to the radio emission and to
calculate its effect on the spectral index. 
Then, we discuss within
a simple model the influence  of energy losses, the type of propagation
and escape of CREs  on the spectral index.

\section{Radio emission of SNRs}

\subsection{Theoretically expected value}

We estimate the fraction of the
nonthermal radio emission of a galaxy that is expected to come
from SNRs in the framework of a simple leaky-box model.
The number of CREs in a certain
energy range $\mathrm{d}E$, $N(E)$, can then be calculated as:
\begin{equation}
\frac{\partial N(E)}{\partial t} + \frac{N(E)}{\tau} = Q(E).
\label{eqone}
\end{equation}
Q(E) is the production rate of CREs and $\tau$ the life-time of CREs.
The main sources of CREs  are gravitational supernovae (SN) 
(SN II and Ib/c)  (e.g. Xu et al. \cite{xu_a} and discussion 
in Xu et al \cite{xu_b}).
The energy spectrum of the CREs produced can be
described by a power law:
\begin{equation}
Q(E)=\nu_\mathrm{SN}\,q_\mathrm{SN}\,\left(\frac{E}{E_0}\right)^{-\gamma},
\label{eqtwo}
\end{equation}
where $\nu_\mathrm{SN}$ is the SN-rate, $q_\mathrm{SN}$  the number of
CREs produced per energy interval per SN and $\gamma$ the energy 
injection spectral index.
For  CREs in
SNRs $\tau$ is given by the life-time of the SNR, $\tau_\mathrm{SNR}$, 
and for the diffuse CREs $\tau=\tau_\mathrm{diff}$ 
can be expressed as:

\begin{equation}
\frac{1}{\tau_\mathrm{diff}}=\frac{1}{\tau_\mathrm{esc}}+
\frac{1}{\tau_\mathrm{loss}},
\label{eqthree}
\end{equation}
where $\tau_\mathrm{esc}$ is the escape time-scale from the galaxy and
$\tau_\mathrm{loss}$ is the energy loss time-scale. 
We take into account a possible energy dependence of  
$\tau_\mathrm{diff}$ according to:

\begin{equation}
\tau_\mathrm{diff}(E)=\tau_\mathrm{diff}(E_0) \left(\frac{E}{E_0}\right)^{-y}.
\label{eqfour}
\end{equation}
The exponent $y$ can have different values depending on the
characteristics of the propagation of CREs.
It is $y=1$ if $\tau_\mathrm{diff}=\tau_\mathrm{loss}$ and synchrotron and 
inverse Compton losses are the dominant energy loss processes.
If $\tau_\mathrm{diff}=\tau_\mathrm{esc}$ and CREs propagate 
mainly by diffusion, 
the energy dependence of the diffusion coefficient,
$D(E)=D_0 E^{-\mu}$, gets reflected in 
$y = \mu \approx 0.5$.
If, on the other hand, CREs can escape freely 
($\tau_\mathrm{diff}=\tau_\mathrm{esc}$)
by convection or if adiabatic
and bremsstrahlung losses are dominant, we expect $y=0$.
 
In the case of a steady state ($\partial N/\partial t=0$) 
the solution of eq. (\ref{eqone}) is simply:

\begin{equation}
N(E)=Q(E) \, \tau.
\label{eqfive}
\end{equation}

We make the simplification that an electron with energy $E$ emits all
the synchrotron radiation at the frequency

\begin{equation}
\nu  =  0.435 \left(\frac{E}{m_\mathrm{e}c^2}\right)\nu_\mathrm{G} 
 =  4.65 \left( \frac{E}{1\mathrm{GeV}}\right)^2
\left(\frac{B}{\mu\mathrm{G}}\right) \mathrm{MHz}
\label{eqsix}
\end{equation}
(Longair \cite{longair}, Vol. 1 p. 332) where $\nu_\mathrm{G}$ is 
the gyrofrequency.
Then the total synchrotron  emission is given by:
\begin{equation}
P(\nu)\mathrm{d}\nu=N(E) \left(\frac{\mathrm{d}E}{\mathrm{d}t}\right)
_\mathrm{syn} 
\left(\frac{\mathrm{d}E}{\mathrm{d}\nu}\right)\mathrm{d}\nu,
\label{eqseven}
\end{equation}
with the energy losses by synchrotron radiation 
\begin{eqnarray}
\left(\frac{\mathrm{d}E}{\mathrm{d}t}\right)_\mathrm{syn} & = &
\frac{4}{3} \sigma_\mathrm{T} c \left(\frac{E}{m_\mathrm{e}c^2}\right)^2
U_\mathrm{B} \nonumber \\
& = & 2.5 \,  10^{-18}\left(\frac{B}{\mu \mathrm G}\right)^2
\left(\frac{E}{\mathrm{GeV}}\right)^2 \mathrm{GeV s^{-1}},
\label{eqeight}
\end{eqnarray}
where $\sigma_\mathrm{T}$ is the Thompson scattering cross section,
$B$ the magnetic field strength and $U_\mathrm{B}$ its energy density.

With the above equations, we can express the ratio of 
the diffuse synchrotron emission to the emission from SNRs as:

\begin{eqnarray}
\frac{P_\mathrm{diff}(\nu)}{P_\mathrm{SNR}(\nu)}  &=&
\frac{N(E)_\mathrm{diff} 
\left(\mathrm{d}E/\mathrm{d}t\right)_\mathrm{syn,diff} 
\left(\mathrm{d}E/\mathrm{d}\nu\right)_\mathrm{diff}}
{N(E)_\mathrm{SNR} \left(\mathrm{d}E/\mathrm{d}t\right)_\mathrm{syn,SNR} 
\left(\mathrm{d}E/\mathrm{d}\nu\right)_\mathrm{SNR}}  \nonumber\\
&=& \frac{E^{-\gamma} \tau_\mathrm{diff} E^2 B_\mathrm{diff}^2
\left(\mathrm{d}E/\mathrm{d}\nu\right)_\mathrm{diff}}
{E^{-\gamma} \tau_\mathrm{SNR} E^2 B_\mathrm{SNR}^2
\left(\mathrm{d}E/\mathrm{d}\nu\right)_\mathrm{SNR}} \nonumber \\
& = & \left(\frac{\nu}{\nu_0}\right)^{-\frac{y}{2}}
\left(\frac{B_\mathrm{diff}}{B_\mathrm{SNR}}\right)^{\frac{\gamma+1}{2}}
\frac{\tau_\mathrm{diff}(\nu_0)}{\tau_\mathrm{SNR}}.
\label{eqnine}
\end{eqnarray}
Thus, the ratio of the diffuse synchrotron emission to the emission from SNRs
is mainly determined by the ratio of the life-times of
CREs in the two zones and by the ratio of the magnetic field 
strengths. 
The magnetic field strength in SNRs 
is expected to be higher than in the diffuse 
ISM because the gas in which the magnetic field is frozen-in gets
compressed by the shock. In general, the magnetic field strength
found in SNRs exceeds the value expected by this compression which
would be a factor of 4 in a strong shock, indicating that the magnetic
field is not only compressed, but enhanced by the shock. 
Especially in young SNRs, the inferred values span
a wide range from a few $\mu$G (SN 1006; Tanimori et al. \cite{tanimori}) 
to much higher values of
70 $\mu$G (Kepler's SNR; Matsui et al. \cite{matsui}), $>$ 80 $\mu$G (Cas A;
Cowsik \& Sarkar \cite{cowsik}) and  even 300 $\mu$G 
(Crab Nebula; Kennel  \& Coroniti \cite{kennel}).
Taking these  values as a guide, we  assume for our present 
estimate  $B_\mathrm{SNR}=75 \mu$G and
$B_\mathrm{diff}=5 \mu$G. 
Furthermore, we adopt for the injection spectral index $\gamma=2.2$ 
(V\"olk et al. \cite{voelk_a}),
for the life-time  of a SNR its adiabatic phase $\tau_\mathrm{SNR}\approx 3\,10^4$ yr
and for $\tau$ the energy loss time scale 
due to synchrotron and inverse Compton losses. The latter  yields for
a magnetic field of 5 $\mu$G and a radiation field of 
energy density 1 eV cm$^{-3}$
$\tau_\mathrm{diff}=2.5 \,10^7$ yr  at $\nu_0=1.5$ GHz. 
With these values we get
$P_\mathrm{SNR}/P_\mathrm{diff}=0.09$  at a frequency 1.5 GHz.

A further contribution to the radio emission of a galaxy 
are radio supernovae (RSNe, e.g. Weiler et al. \cite{weiler}). 
Their contribution is difficult to 
estimate due to uncertainties in their luminosity and life-time. 
P\'erez-Olea \& Colina (\cite{perez}) have included RSNe in a 
model for the radio emission of starburst galaxies by generalizing
and extrapolating the luminosities of 9 observed RSNe. 
Adopting their numbers, we calculate that for a galaxy in a steady state  
the total radio emission of RSNe is, for a life-time of a RSNe of 100 yr,
$1.5 \times$ the  total radio emission of SNRs. For
a shorter life-time of 10 yr, their total radio emission
would be  a factor of 2 lower.
Thus, in spite of the uncertainties,
it seems plausible that RSNe contribute at a level  of at least
a few percent  to the total radio emission of a galaxy.

\subsection{Comparison to observations}

In order to compare this theoretically derived ratio to 
observations we calculate the expected average radio emission
of a SNR:
\begin{eqnarray}
& & P_\mathrm{SNR}(\nu)\mathrm{d}\nu =  q_\mathrm{SN}\,
\left(\frac{E}{E_0}\right)^{-\gamma}
\left(\frac{\mathrm{d}E}{\mathrm{d}t}\right)_\mathrm{syn} 
\left(\frac{\mathrm{d}E}{\mathrm{d}\nu}\right) \mathrm{d}\nu\nonumber \\
& = &  1.07 \, 10^{-25} 
\left(\frac{q_\mathrm{SN}}{\mathrm{eV}^{-1}}\right) 
\left( \frac{\nu}{\mathrm{MHz}} \right)^{\frac{1-\gamma}{2}}
\left( \frac{B}{\mu \mathrm{G}} \right)^{\frac{1+\gamma}{2}} 
\mathrm{W Hz^{-1}}
\label{eqten}
\end{eqnarray}
where we have inserted  the relations from eqs. (\ref{eqsix}) and 
(\ref{eqeight}).
For $\gamma=2.2$ this predicts a  radio spectral index of SNRs of 0.6,
in reasonable agreement with the observations.
The total number of CREs per energy interval
produced by a SN, $q_\mathrm{SN}$,  can be estimated
from the total energy released by a SN which we take as $10^{51}$ erg.
Assuming that 10\% of this energy goes into CR acceleration  and of this
1\% into the electron component,
a total energy of $E^\mathrm{e}_\mathrm{SN}=10^{48}$erg is available per SN to
produce the relativistic electrons following a power-law spectrum:
\begin{equation}
E^\mathrm{e}_\mathrm{SN}=q_\mathrm{SN}\int^\infty_\mathrm{E_{min}} 
E \left(\frac{E}{E_0}
\right)^{-\gamma} \mathrm{d}E.
\label{eqeleven}
\end{equation}
With $\gamma=2.2$, assuming for the lower energy limit of the
produced spectrum $E_\mathrm{min}=1$ GeV and adopting the
scaling parameter $E_0=1$ GeV this yields
\begin{equation}
q_\mathrm{SN}=\frac{E^\mathrm{e}_\mathrm{SN}}{E_0^2}\left(\frac{E_\mathrm{min}}
{E_0}\right)^{\gamma-2} (\gamma-2)=1.25\times 10^{41} \mathrm{eV^{-1}}.
\label{eqtwelve}
\end{equation}

In total, we obtain for the radio emission of a SNR, 
taking $B_\mathrm{SNR}=75 \mu$G,
\begin{equation}
P_\mathrm{SNR}= 2.1\, 10^{17}  \left(\frac{\nu}{\mathrm{GHz}} \right)
^{-0.6} \mathrm{W Hz^{-1}}
\label{eqthirteen}
\end{equation}
This can be compared to the average observed emission of a SNR which
can be derived from the relation between the surface brightness 
$\Sigma$ and the diameter $D$ of SNRs. 
We adopt the relation found by Huang et al. (\cite{huang})
who fitted SNRs from the Galaxy, \object{M~82} and the 
\object{Magellanic Clouds}:
\begin{equation}
\Sigma_{8.4 \mathrm{GHz}}=6.6\, 10^{-16} \left(\frac{D}
{\mathrm{pc}}\right)^{-3.6}
\mathrm {W m^{-2} Hz^{-1} sr^{-1}}.
\label{eqfourteen}
\end{equation}

Following Condon \& Yin (\cite{condon})
one can connect the diameter and the age of the radio emitting SNR
by:
\begin{equation}
\left(\frac{D}{\mathrm{pc}}\right)=0.43 \left(\frac{E_{50}}{n}\right)
\left(\frac{t}{\mathrm{yr}}\right)^{\frac{2}{5}},
\label{eqfifteen}
\end{equation}
where $E_{50}$ is the energy of a SN going into particle acceleration
in units of $10^{50}$ erg and $n$ is the gas density in particles
per cm$^{3}$. In the following we assume $E_{50} = 1$ and $n=1$.
Using $L=4 \pi^2 (D/2)^2 \Sigma$ we then calculate the 
average luminosity of a SNR
by integrating the 
luminosity evolution from $t=0$ to $t=\tau_\mathrm{SNR}=3\,10^4$ yr
and adopting the  frequency dependence of eq. (\ref{eqthirteen}), 
$P_\mathrm{SNR}\propto \nu^{-0.6}$:
\begin{eqnarray}
& &<P_\mathrm{SNR}>  =  4 \pi^2 \left(\frac{D}{2}\right)^2 
\left(\frac{1}{\tau_\mathrm{SNR}}\right)  \nonumber \\
& & \int_0^{3\,10^4}\Sigma(D(t)) = 3.3 \, 10^{17} 
\left(\frac{\nu}{\mathrm{GHz}}\right)^{-0.6} \mathrm{W Hz^{-1}},
\label{eqsixteen}
\end{eqnarray}
which is very close to the above theoretical value (eq. 13).

We can now estimate the total emission of SNRs expected
for our Galaxy:
Adopting a SN rate of 1 SN per 30 yr (Berkhuijsen \cite{berkhuijsen}), 
the total number of the SNRs present in
our Galaxy is, with $\tau_\mathrm{SNR}=3\,10^4$ yr,
$N_\mathrm{SNR}=\nu_\mathrm{SN}\tau_\mathrm{SN}=1000$. 
This yields, together with eq. (\ref{eqthirteen}), 
a total
radio emission from SNRs of $P_\mathrm{SNR,tot}(1.5\, \mathrm{GHz}) =
P_\mathrm{SNR}(1.5\,\mathrm{GHz}) N_\mathrm{SNR}=2.6\, 10^{20} 
\mathrm{W Hz^{-1}}$.
This can be compared to   the total radio emission of our Galaxy
which is at 408 MHz $8.7\,10^{21} \mathrm{W Hz^{-1}}$ 
(Beuermann et al. \cite{beuermann}). 
Extrapolating this value, with a radio spectral
index of 0.8, to 1.5 GHz we get  
$3\,10^{21}$ W Hz$^{-1}$. From this calculation we derive
a contribution of 
about 9\% from SNRs to the total radio emission, in
satisfactory agreement with  the value derived in the previous
section in spite of the crudeness of both estimates. 
Therefore, we conclude that  $\simeq$ 10\% is
a reasonable estimate  for the contribution of SNRs to the total
radio emission. This value is also in agreement with previous 
estimates by other authors, who
compared the expected total synchrotron emission from SNRs with
the observed radio emission of galaxies, and have found that
SNRs alone can only explain  about 10\% of the
total radio emission (Biermann \cite{biermann}; Ulvestad \cite{ulvestad}). 

\subsection{Influence on the spatially integrated radio spectral index}

\begin{figure}
\resizebox{\hsize}{!}{\includegraphics{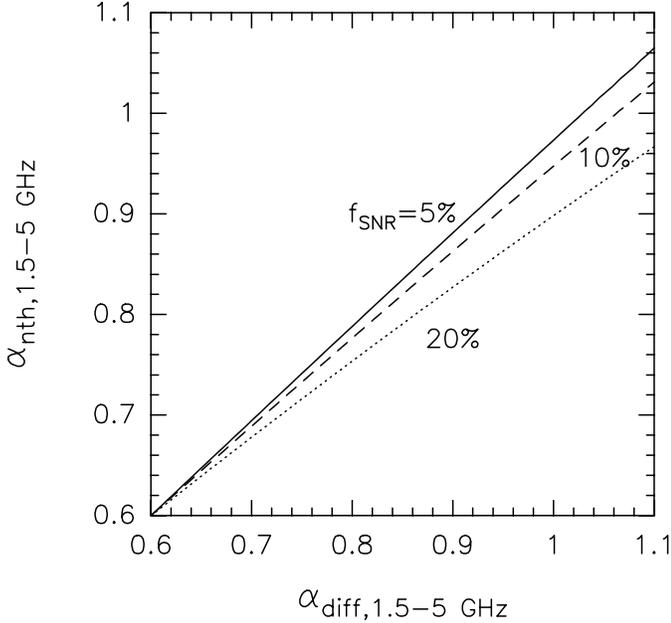}}
\caption{The nonthermal spectral index,
$\alpha_\mathrm{nth}$, as a function of the diffuse radio spectral index,
$\alpha_\mathrm{diff}$,
between 1.5 and 5 GHz, for different contributions of the radio
emission from SNRs to the radio emission: 
$f_\mathrm{SNR}=$5\% (full line), 10\% (dashed 
line), 20\% (dotted line).}
\label{figone}
\end{figure}

In Fig.~\ref{figone} we show the total (i.e. spatially integrated) 
nonthermal spectral index, $\alpha_\mathrm{nth}$,
as a function of the spectral index of the diffuse
radio emission, $\alpha_\mathrm{diff}$, 
for different contributions of SNRs to
the radio emission, expressed by 
$f_\mathrm{SNR}=P_\mathrm{SNR}/P_\mathrm{nth}$. We assume, as above,
an (energy) injection spectral index for CREs of $\gamma=2.2$, yielding
a (frequency) spectral index of SNRs of 
$\alpha_\mathrm{SNR}=(\gamma-1)/2=0.6$ (eq. (\ref{eqten})). A
lower source spectral index of $\gamma = 2.0$ to $2.1$ is indicated by
recent calculations by Berezhko and V\"olk (\cite{berezhko}). This would give
$\alpha_\mathrm{SNR} = 0.5$ to $0.55$ and would somewhat 
increase the difference
between $\alpha_\mathrm{nth}$ and $\alpha_\mathrm{diff}$ for a
given $f_\mathrm{SNR}$.

The radio emission of SNRs has a noticeable effect on the
nonthermal radio spectral index.  
Due to the contribution of SNRs it gets flatter, especially
for large values of $\alpha_\mathrm{diff}$ where  the spectral index is
lowered by $\sim 0.1$ for  $f_\mathrm{SNR}=0.1$.

\section{CR propagation and energy losses}


The types of propagation of CREs in a galaxy are diffusion,
i.e. the CREs are scattered randomly by magnetic field irregularities,
and convection, a systematic movement of the scattering centres
outwardly. While propagating through a galaxy, CREs 
suffer energy losses.
Eventually, CREs might escape from the galaxy. 

In order to discuss the radio spectral index in the various cases,
we calculate two simple models, one that considers
diffusion and one in which we study convection. 
In both models we take into account
energy losses by synchrotron and inverse Compton losses.
Here, we ignore adiabatic and bremsstrahlung losses which 
might also be important,
especially in starburst galaxies, but since they do not change
the spectral index, they are in the framework of the present 
study equivalent to  no energy losses at all.

\subsection{Diffusion model}

The diffusion model 
can be described by the following equation:

\begin{equation}
D(E)\frac{\partial^2 N(E,z)}{\partial z^2} + \frac{\partial}{\partial E}
\left(b(E) N(E,z)\right) = - Q(E,z),
\label{eqseventeen}
\end{equation}
where $z$ is the height above the galactic midplane. We assume that the
sources of CRs are situated in the midplane at $z=0$ and that their
energy dependence is given by a power-law according to
eq. (\ref{eqtwo}). The synchrotron and inverse
Compton losses, $b(E)$, are assumed to be spatially homogeneous 
and are given by:

\begin{equation}
b(E)= - \frac{\mathrm{d}E}{\mathrm{d}t} \bigg|_\mathrm{syn+iC} = \frac{4}{3} 
\sigma_\mathrm{T} c
\left(\frac{E}{m_\mathrm{e} c^2} \right)^2 (U_\mathrm{rad}+U_{B}).
\label{eqeighteen}
\end{equation}
$U_\mathrm{rad}$ is the energy density of the radiation field below the
Klein-Nishina limit (e.g. Longair \cite{longair}). 
We adopt a free escape boundary at $|z| = z_\mathrm{h}$, the outer boundary of the
halo, so that $N(E,|z_\mathrm{h}|)=0$. An approximate solution is given by
(for further details see Lisenfeld et al. \cite{lisenfeld_a}):

\begin{eqnarray}
& & N(E,z)   =  \nu_\mathrm{SN}
\left({E\over m_\mathrm{e} c^2}\right)^{-\gamma}
\left({E \over 1\mathrm{GeV}}\right)^{-\mu}  \nonumber \\
& &\left({{_{1}F_1}(p+{1\over 2},{3\over 2},s_\mathrm{h}) 
\over {_{1}F_1}(p,{1\over 2},s_\mathrm{h}) }
   {_{1}F_1}(p,{1\over 2},s) \cdot |z_\mathrm{h}|  \right.\nonumber \\
& & -\left. {_{1}F_1}(p+{1\over 2},{3\over 2},s) \cdot |z|\right)
\label{eqnineteen}
\end{eqnarray}
with
\begin{equation}
p=-\left({2x+\mu-3 \over 2(1-\mu)}\right),
\label{eqtwenty}
\end{equation}
${_{1}F_1}(a,b,x)$ denoting the confluent hypergeometric function
(or Kummer function), and 
\begin{equation}
s= -{c_1 (U_\mathrm{rad}+U_\mathrm{B}) (1-\mu)\over 4 D_0 } z^{2}
\left({E\over m_\mathrm{e} c^2}\right)
\left({E \over 1\mathrm{GeV}}\right)^{-\mu}
\label{eqs}
\end{equation}
and
\begin{equation}
s_\mathrm{h}=-{ c_1(U_\mathrm{rad}+U_\mathrm{B})  (1-\mu)\over 4 D_0 }
 z_\mathrm{h}^{2} \left({E\over m_\mathrm{e} c^2}\right)
\left({E \over 1\mathrm{GeV}}\right)^{-\mu}
\label{eqsh}
\end{equation}
where
\begin{equation}
c_1={4\over 3} {\sigma_{\mathrm T} c\over m_\mathrm{e} c^2}.
\label{eqcone}
\end{equation}

\subsection{Convection model}

The convection model is described by the following equation:

\begin{eqnarray}
Q(E,z) & = & \frac{\partial}{\partial z} \left(V N(E,z)\right) \nonumber \\
& - & \frac{\partial}{\partial E} \left\{\left( \frac{1}{3} 
\frac{\partial V}{\partial z} E + b(E) \right) N(E,z) \right\} 
\label{eqconv}
\end{eqnarray}
where $V$ is the convection speed, assumed here to be constant. The solution
of this equation can be obtained with the method of characteristics:

\begin{equation}
N(E,z)= \frac{q}{2 V} E^{-\gamma} \left(1-\frac{b E z}{V} \right)^{\gamma-2}.
\label{eqconvsolv}
\end{equation}

\subsection{Asymptotic spectral indices}

The diffuse synchrotron emission is calculated from $N(E,z)$ in the 
same way as for SNRs (eqs. (\ref{eqseven}) and (\ref{eqeight})).
%
%
%
The total diffuse synchrotron emission
is obtained by integrating $P_\mathrm{diff}$ over the extent of the halo,
$z_\mathrm{halo}$,
which in the case of the diffusion model coincides with the 
free-escape boundary, $z_\mathrm{h}$.

We define the escape probability, $P_\mathrm{esc}(\nu)$,
as the ratio of the measured total nonthermal radio emission
(including the diffuse synchrotron emission and the radio
emission  from SNRs),
$P_\mathrm{nth}(\nu)$, to the maximum possible nonthermal radio emission,
$P_\mathrm{nth,max}(\nu)$, that would be measured if the halo were infinitely
large.

The expected diffuse and nonthermal  
spectral indices  for different asymptotic cases are:

Case (1): Synchrotron and inverse Compton 
losses are very strong so that no escape is possible. Then, independent
of whether diffusion or convection is the principle mode of
propagation, we have  $\alpha_\mathrm{diff}=
\gamma/2=1.1$. Including  the contribution of SNRs, the
spectral index becomes (see Fig.~\ref{figone})  
$\alpha_\mathrm{nth} \simeq 1.0$. For
a lower injection spectral index, $\gamma = 2.0$,
we would get $\alpha_\mathrm{nth} \simeq 0.9$.

Case (2): CREs can escape freely from the galaxy 
by diffusion without suffering synchrotron and inverse Compton 
losses. The diffusion coefficient 
can be energy dependent, $D(E)\propto E^{\mu}$.
For our Galaxy, $\mu$ has been determined from the
energy-dependence of the fraction of secondary to primary
CR nuclei. For CRs with energies above a few GeV, the energy range
relevant for radio observations in the GHz range, 
$\mu = 0.4 -0.7$ has been found (Garc\'\i a-Mu\~noz et al. \cite{garcia}). 
The expected diffuse spectral index is then, adopting $\mu=0.5$,
$ \alpha_\mathrm{diff}=(\gamma + \mu -1)/2 = 0.85$, 
and including  the contribution of SNRs it becomes
$\alpha_\mathrm{nth}\simeq 0.82$, with corresponding 
reductions for $\gamma = 2.0$.

Case (3):  CREs propagate by convection
 and do not suffer
considerable synchrotron or inverse Compton losses.  In this case we expect
$\alpha_\mathrm{diff}=\alpha_\mathrm{nth}=(\gamma-1)/2 = 0.6$ 
for $\gamma = 2.2$;
the spectral index remains unchanged when including the emission from
SNRs. 

\subsection{The spatially integrated spectral index}

\begin{figure}
\resizebox{\hsize}{!}{\includegraphics{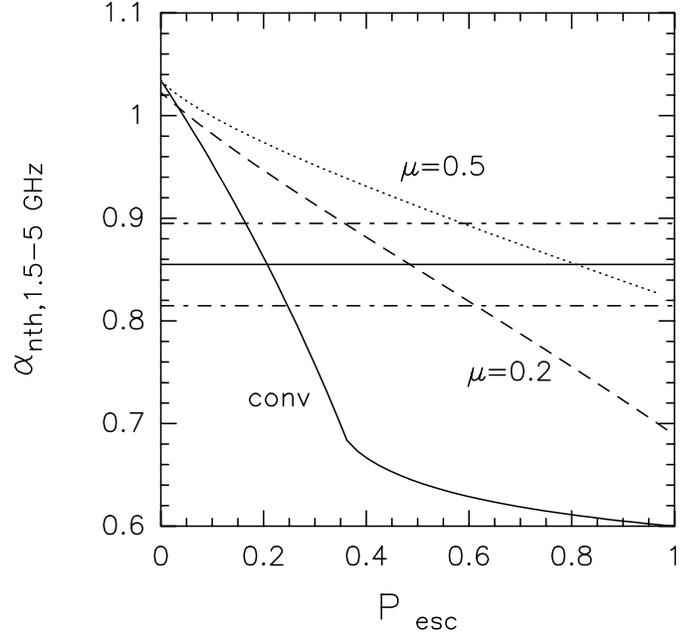}}
\caption{The nonthermal spectral index between 1.5 and 5 GHz,
assuming $\gamma = 2.2$ and $f_\mathrm{SNR}=0.1$, is shown as a function of the
escape probability
$P_\mathrm{esc}$. The full line is calculated for a convection model,
the dashed line for a diffusion model with $\mu=0.2$ and the
dotted line for a diffusion model with $\mu=0.5$. 
The full and dashed-dotted horizontal lines give the observational
value for $\alpha_\mathrm{nth}$ and its dispersion, calculated as the
average of the results by Niklas et al. (\cite{niklas_b}) and 
Klein (\cite{klein}). For $\gamma = 2.0$ all model 
curves would be lower by about 0.1.}
\label{figtwo}
\end{figure}

The results of the models are shown in Fig.~\ref{figtwo} 
where the nonthermal spectral
index (assuming a contribution of SNRs of 10\% at 1.5 GHz)
is plotted as a function of the escape probability for the case of 
diffusion and of convection.
The values for $\alpha_\mathrm{nth}$ predicted by the two models are
compared to the observed mean nonthermal spectral index 
which is about $0.85 \pm 0.04$.
This observational value  is obtained as the
average of the results from results by Klein (\cite{klein}) ($0.88 \pm 0.06$)
and Niklas et al. (\cite{niklas_b}) ($0.83 \pm 0.02$)
who determined the average $\alpha_\mathrm{nth}$ of samples of galaxies 
by subtracting the thermal radio emission from the observed total
radio emission. 

Within the pure diffusion model with $\mu=0.5$ 
the observed $\alpha_\mathrm{nth}$ is explained by a high escape probability 
($P_\mathrm{esc} \stackrel{>}{_{\sim}} 50\%$). For a weaker energy dependence 
of the diffusion coefficient, $\mu=0.2$, however, the predicted escape
probability is lower,
$P_\mathrm{esc}=30-60\%$. 
The conclusion from the convection model
is very different: Here the observed $\alpha_\mathrm{nth}$ is 
explained by $P_\mathrm{esc}\simeq 25\%$. The different conclusions with 
respect to  the
escape probability given by the diffusion and the convection model
are due to the fact that (i) convection is an energy independent 
process and (ii) in the case of convection the relation between
the spectral index and the escape probability is not linear, in the sense
that low escape probabilities are predicted for almost the whole range
of spectral indices.

We conclude that from the observed average radio spectral
index of galaxies it can neither be decided 
to which degree CREs can escape from galaxies
nor whether diffusion or convection is the dominant mode of propagation.
Only for asymptotic cases partial conclusions can be drawn: If
the spectral index of a galaxy is very steep ($\stackrel{>}{_{\sim}}0.9$)  
energy losses
are important and the escape rate is low. Yet, it cannot be
determined whether diffusion or convection is the dominant type
of propagation. 
A very flat overall spectral index ($\stackrel{<}{_{\sim}} 0.7$),
on the other hand, indicates  strong convection or almost energy independent
diffusion. Since measurements in our Galaxy indicate that diffusion
does depend significantly on energy ($\mu = 0.4-0.7$, Garc\'\i a-Mu\~noz et al.
\cite{garcia}), convection is the more likely explanation.  In
this case, however, we cannot
draw any conclusions with respect to the confinement of CREs.

\subsection{Spatially resolved spectral index}

\begin{figure}
\resizebox{\hsize}{!}{\includegraphics{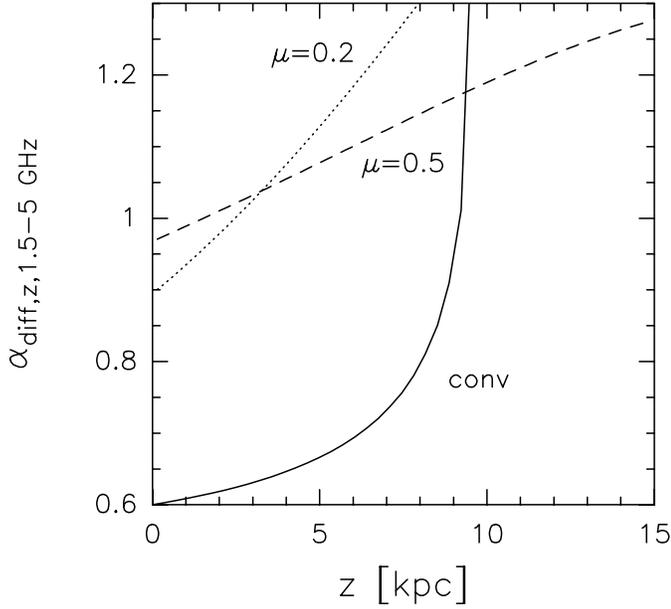}}
\caption{The diffuse radio spectral index as a function of the 
height over the galactic plane, $z$, in the case of
important energy losses. 
The full line is calculated for a convection model,
the dashed line for a diffusion model with $\mu=0.5$ and the dotted
line for a diffusion model with $\mu=0.2$.
We have assumed a magnetic field $B=5\, \mu$G, a radiation field of energy 
density 1 eV cm$^{-3}$, convection velocity
$V=200$ km s$^{-1}$ and diffusion coefficient $D_0=10^{29}$ cm$^2$ s$^{-1}$.} 
\label{figthree}
\end{figure}
\begin{figure}
\resizebox{\hsize}{!}{\includegraphics{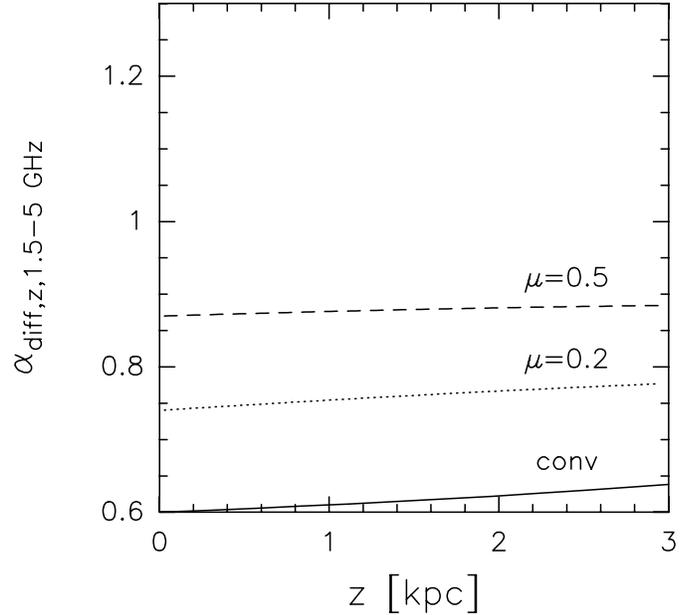}}
\caption{The diffuse radio spectral index as a function of the 
height over the galactic plane, $z$,  in the case of almost
free escape. 
The line coding and parameters are as in Fig.~\ref{figthree}.}
\label{figfour}
\end{figure}

The question of how important energy losses are for the propagation of
CREs in a galaxy can be decided if the spectral index distribution in the
halo is known. If the spectrum steepens considerably away from the
disk, it is a clear sign that CREs suffer considerable synchrotron
and inverse Compton losses as they propagate outwardly.
 This is illustrated in Fig.~\ref{figthree} and \ref{figfour}, 
where we show the diffuse radio spectral index, 
$\alpha_{\rm diff,z}$, calculated
from the convection and diffusion models of Section 3.1,
as a function of the distance from the galactic plane, $z$.
In Fig.~\ref{figthree} we show the results for a large halo 
($z_h =15$ kpc) in which the
energy losses affect the CREs considerably before they manage to escape.
The escape probabilities are here 9\% for convection  and 
6\% ($\mu=0.2$), respectively 10\% ($\mu=0.5$) for diffusion.
In Fig.~\ref{figfour} the opposite case is shown: A small halo 
($z_h =3$ kpc) where
the CREs can escape without suffering significant energy losses, so that
the escape probabilities lie between 78\% for convection
and 85\% ($\mu=0.2$), respectively 87\% ($\mu=0.5$) for diffusion.
We find that important  energy losses are reflected clearly in 
a steepening of the spectral index with increasing distance from the
plane. On the contrary, if the CREs can escape almost freely,
or if only adiabatic or bremsstrahlung losses occur,
no such steepening is seen.

\subsection{Comparison to observations: Are energy losses important?}

\begin{figure}
\resizebox{\hsize}{!}{\includegraphics{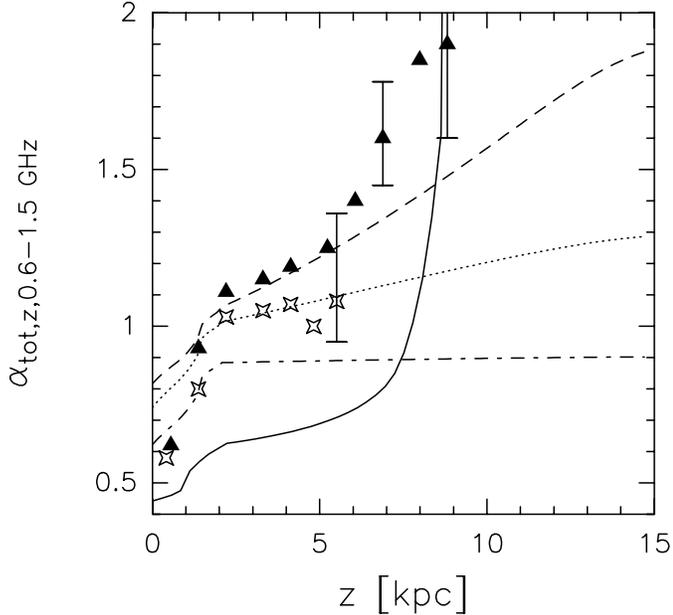}}
\caption{A comparison of our model results to the data of NGC 891
(Hummel et al. \cite{hummel_c}). The triangle refer to the spectral
index in the western part of the halo, and the stars to the 
eastern side of the halo. The full line corresponds to the convection
model and the other lines to the diffusion model with $\mu=0.5$ and
different values for the diffusion coefficient and thus escape
probability (see text).}
\label{figfive}
\end{figure}
The distribution of the spectral
index in the halo is a definite diagnostic of whether energy losses
are important. The observational study of 
radio halos in external galaxies is difficult, because the halos
are extended and intrinsically faint (see Dahlem \cite{dahlem_b}).
Therefore, not many galaxies with multi-frequency
radio data for the halo are known. 

One of the best studied radio
halos is that of the spiral galaxy \object{NGC~891}  
(Hummel et al. \cite{hummel_c}). Its spatially integrated nonthermal
radio spectral index is 0.78 (Niklas et al. \cite{niklas_b}),
only slightly below the average spectral index of galaxies
(see Fig~\ref{figtwo}).
In Fig.~\ref{figfive} we show its radio spectral index distribution in 
the halo. The data are at a resolution of 40'', corresponding
to 1.8 kpc at the distance of 9.5 kpc of \object{NGC~891}. 
The spectrum  
steepens very quickly within the first 2-3 kpc.
It continues to steepen more slowly outside this range,
especially in the eastern side of the halo. Here, 
the steepening becomes more pronounced at  
$z \stackrel{>}{_{\sim}} 6$ kpc.

In Fig. ~\ref{figfive} we also show the results of our model in
which we have included, in order to correspond to the data, 
apart from the diffuse radio emission,
the contribution from SNRs and the thermal radio emission.
For the latter two components we have 
assumed, $f_{\rm SNR}=0.1$, as before,
and a contribution of the thermal radio emission of 10 \% at 1.5 GHz,
a typical value for normal galaxies (Condon 1992). 
We furthermore assume
that the thermal radio emission and the radio emission from SNRs is
restricted within 100 pc above the disk (the exact value of 
this spatial extent is not important  because of the low 
resolution of the data).
Finally, we convolve the model results with a Gaussian beam of 
HPBW 1.8 kpc, corresponding to the resolution of the data.

We have adjusted the diffusion coefficient and convection velocity
such that we achieved the best possible fit to the data.
Because of the simplicity of the models described here, 
we cannot expect to describe the data in all detail, we are
aiming rather at a qualitative  comparison with  the main 
features. For a quantitative comparison
a more complex model would be necessary, combining diffusion and 
convection, taking into
account spatial variations of the parameters ($B, D_0$, etc.)
and describing more realistically the distribution of SNRs and
the contribution  
of the thermal radio emission. Nevertheless, the diffusion model
is able to describe a large range of the data reasonably well.
The fast steepening of the spectral index within the first
2-3 kpc above the disk can to a large extent
be attributed to the contribution of SNRs and the thermal
radio emission close to the disk.
The steepening outside this range requires the presence
of dominant energy losses: The diffusion models that
describe the data well have escape probabilities of
11 \% (dotted line) and 0 \% (dashed line). 
For comparison a model with a high escape rate
is shown (78 \% -- dashed-dotted line) which is not able to
describe the data.
In order to reproduce the fast steepening of the spectral index
at $z \stackrel{>}{_{\sim}} 6$ kpc
in the western part of the halo, we would possibly have
combine the diffusion model
with the convection model. 

Radio halos have been observed in several other galaxies, 
although mostly not in such detail.
In all of these galaxies
a steepening of the spectral index  
at least up to $\alpha_\mathrm{nth}\stackrel{>}{_{\sim}} 1.0$
is found with increasing distance from the disk.
This is the case for both starburst galaxies, 
as \object{M~82} (Seaquist \& Odegard \cite{seaquist}; 
Reuter et al. \cite{reuter});
\object{NGC~253} (Carilli et al. \cite{carilli}); 
\object{NGC~2146} (Lisenfeld et al. \cite{lisenfeld_b}); 
\object{NGC~4666} (Sukumar et al. \cite{sukumar}; Dahlem et al. 
\cite{dahlem_a})
and non-starburst spiral galaxies like 
\object{NGC~4631} (Hummel \& Dettmar \cite{hummel_b}); 
\object{NGC~5775} (Duric et al. \cite{duric});
\object{NGC~5055} and \object{NGC~7331} (Hummel \& Bosma \cite{hummel_a}).
We conclude therefore that for all galaxies for which a 
multi-frequency radio halo has been observed,
energy losses are important and it seems likely that this is
generally the case in galactic
halos.

Niklas \& Beck (\cite{niklas_a}) have found a trend that actively star-forming
galaxies tend to have lower spectral indices and more quiescent 
galaxies steeper ones. They interpret this as an indication
that in galaxies with a high star-formation efficiency (SFE) 
CREs escape more easily due to a 
galactic wind. On the basis of the above consideration, however,
it seems more likely that in these galaxies
convection causes the flat spatially integrated spectrum.
Energy losses play an important role also in galaxies with
a high SFE  which is shown by the steepening of the spectral index
in the halo. A good example is the starburst galaxy  \object{M~82}.
For this galaxy the observed overall synchrotron spectral 
index $\alpha_\mathrm{nth} = 0.66$
(Niklas et al. \cite{niklas_b}) 
indicates strong convection (see Fig. 2) whereas at the same time
the steepening of the spectral index in the halo shows the presence
of important energy losses (Seaquist \& Odegard \cite{seaquist}; 
Reuter et al. \cite{reuter}).

\subsection{Consequences for the FIR/radio correlation}

The question whether CREs can escape freely from galaxies or whether
synchrotron and inverse Compton losses determine their spectra
has been asked in the framework of the FIR/radio correlation.
V\"olk (\cite{voelk_b}) has proposed a calorimeter model in which the
tightness of the correlation can be explained by the
fact that CREs loose their energy by synchrotron and inverse
Compton losses below the energy level corresponding to the
observing frequency before they are able to escape from a galaxy.
Lisenfeld et al. (\cite{lisenfeld_a}) have generalized this model, allowing for
moderate escape and a spatially inhomogeneous magnetic field. 
Other authors have claimed the opposite, i.e. that CREs can
escape more or less freely from the galaxy (Chi \& Wolfendale \cite{chi};
Helou \& Bicay \cite{helou}; Niklas \& Beck \cite{niklas_a}).

It has  been argued (Niklas \& Beck \cite{niklas_a}) that the observed
spectral index of galaxies, $\alpha_\mathrm{nth}\simeq 0.85$, 
is in contradiction with the calorimeter
model which predicts in its asymptotic version $\alpha_\mathrm{nth} \simeq
1.0-1.1$, corresponding to case (1) of Section 3.3.
The difference between the prediction of the calorimeter
model and observations of the radio spectral index can be 
decreased by allowing for  moderate escape and a spatially inhomogeneous
magnetic field (Lisenfeld et al. \cite{lisenfeld_a}) 
and even more by including the contribution
of SNRs (this work), so that such a 
modified calorimeter model would predict
$\alpha_\mathrm{nth} \simeq
0.85-0.95$, corresponding to the observed average value. 
However, even in this case this model is not able to account for the 
flatter spectra that some of the galaxies (e.g. \object{M~82}) show
because it is based on diffusion.
In this paper, we have shown, that for galaxies with such a low
spectral index, convection is likely to be  important.
With respect to the question of whether
confinement or escape is taking place the 
overall spectral index is not a good diagnostic.  
Spatially resolved observations
are necessary and these indicate that energy losses indeed do play
a dominant role. Thus, even for galaxies with a low overall spectral
index, like \object{M~82}, escape seems to be negligible as predicted by the
calorimeter model.

\section{Conclusions and summary}

We have estimated and discussed different processes and components
that shape the radio spectral index of a galaxy. In particular,
we have considered the contribution of SNRs, the influence of
different types  of CR propagation (diffusion and convection)
and of energy losses (inverse Compton and synchrotron). Our conclusions
can be summarized as follows:

(i) We have estimated that the radio emission of 
SNRs represents about 10\% of the nonthermal
radio emission of a galaxy. This moderate contribution has a noticeable
effect on the nonthermal radio spectral index, lowering it 
by 0.1 for steep spectra. 

(ii) The spatially integrated radio spectral index of a galaxy is difficult to 
interpret, as it is influenced by various processes. Only for 
asymptotic cases conclusions can be drawn: In galaxies with
very steep spectral indices ($\stackrel{>}{_{\sim}}0.9$) energy losses are high
and the escape rate of CRE is low. A very flat
spectral index ($\stackrel{<}{_{\sim}} 0.7$), 
on the other hand,  is a sign for 
strong convection.

(iii) In order to draw more general conclusions about the CR propagation
and the importance of energy losses, spatially resolved radio 
observations of galactic halos are necessary. A steepening of 
the spectral index away from the galactic disk 
is a clear indication that inverse Compton and
synchrotron losses are important. 

(iv) In all galaxies with multi-frequency radio data for the halo
such a steepening has been  found and therefore 
in these galaxies synchrotron and inverse Compton losses
take place. Thus, it seems likely
that these energy losses are generally important in galactic halos
and that escape rates are low.
The low overall spectral index that is frequently found in starburst
galaxies, at the same time as the
steepening of the spectrum away from the disk,
can be interpreted as convection.
 
\begin{acknowledgements}
We would like to thank the referee, Dr. G. Henri, and M. Dahlem for 
useful comments  on
the manuscript.  
\end{acknowledgements}


\begin{thebibliography}{}

\bibitem[1997]{berezhko} Berezhko, E.G., V\"olk, H.J., 1997, APh 7, 183

\bibitem[1984]{berkhuijsen} Berkhuijsen, E.M., 1984, A\&A 140, 431

\bibitem[1985]{beuermann} Beuermann, K., 
Kanbach, G., Berkhuijsen, E.M., 1985, A\&A 153, 17

\bibitem[1976]{biermann} Biermann, P., 1976, A\&A 53, 295


\bibitem[1992]{carilli} Carilli, C.L., Holdaway, M.A., 
Ho, P.T., De Pree, C.G., 1992, ApJ 399, L59


\bibitem[1990]{chi} Chi, X., Wolfendale, A.W., 1990, MNRAS 245, 101

- Condon, J.J., 1992, ARAA 30, 575

\bibitem[1990]{condon} Condon, J.J., Yin, Q.F., 1990, ApJ 357, 97
                              
\bibitem[1980]{cowsik} Cowsik, R., Sarkar, S., 1980, MNRAS 191, 855

\bibitem[1997]{dahlem_a} Dahlem, M., Petr, M.G., 
Lehnert, M.D., Heckman, T.M., Ehle, M., 1997, A\&A 320, 731

\bibitem[1997]{dahlem_b} Dahlem, M, 1997, PASP 109, 1298

\bibitem[1998]{duric} Duric, N., Irwin, J., Bloemen, H., 1998, A\&A 331, 428

\bibitem[1987]{garcia} Garc\'\i a-Mu\~noz, M., Simpson, J.A., 
Guzik, T.G., Wefel, J.P., Margolis, S.H., 1987, ApJS 64, 269

\bibitem[1998]{green} Green, D.A., 1998, 
`A Catalogue of Galactic Supernova Remnants (1998 September version)', 
Mullard Radio Astronomy Observatory,
Cambridge, United Kingdom 
(available on the World-Wide-Web at 
"http://www.mrao.cam.ac.uk/surveys/snrs/"). 


\bibitem[1993]{helou} Helou, G., Bicay, M.D., 1993, ApJ 415, 93


\bibitem[1994]{huang} Huang, Z.P., Thuan, T.X., Chevalier, R.A., 
Condon, J.J., Yin, Q.F., 1994, ApJ 424, 114

\bibitem[1982]{hummel_a} Hummel, E., Bosma, A., 1982, AJ 97, 242

\bibitem[1990]{hummel_b} Hummel, E., Dettmar, R.-J., 1990, A\&A 236, 33

\bibitem[1991]{hummel_c} Hummel, E., Dahlem, M., 
van der Hulst, J.M., Sukumar, S., 1991, A\& A 246, 10  



\bibitem[1984]{kennel} Kennel, C.F., Coroniti, F.V., 1984, ApJ 283, 710

\bibitem[1988]{klein} Klein, U., 1988, Habilitationsschrift, Universit\"at Bonn

\bibitem[1996a]{lisenfeld_a} Lisenfeld, U., V\"olk, H.J., Xu, C., 1996, 
A\&A 306, 677

\bibitem[1996b]{lisenfeld_b} Lisenfeld, U., Alexander, P., Pooley, G.G.,
Wilding, T., 1996, MNRAS 281, 301

\bibitem[1992]{longair} Longair, M., 1992, High Energy Astrophysics, 
Cambridge University Press

\bibitem[1984]{matsui} Matsui, Y., Long, S.K., Dickel, J.R., 
Greisen, E.R., 1984, ApJ 287, 295

\bibitem[1997]{niklas_a} Niklas, S., Beck, R., 1997, A\&A 320, 54

\bibitem[1997]{niklas_b} Niklas, S., Klein, U., 
Wielebinski, R., 1997, A\&A 322, 19 

\bibitem[1995]{perez} P\'erez-Olea, D. E., Colina, L., 1995, MNRAS 277, 857

\bibitem[1992]{reuter} Reuter, H.P., Klein, U., Lesch, H. 
Wielebinski, R., Kronberg, P.P., 1992, A\&A 256, 10

\bibitem[1991]{seaquist} Seaquist E.R., Odegard, N., 1991, ApJ 369, 320

\bibitem[1988]{sukumar} Sukumar, S., Velusamy, T., Klein, U., 
1988, MNRAS 231, 765

\bibitem[1998]{tanimori} Tanimori T., Hayami, S., Kamei, S., et al., 
1998, ApJ 497, L25

\bibitem[1982]{ulvestad} Ulvestad, J.S., 1982, ApJ 259, 96

\bibitem[1988]{voelk_a} V\"olk, H.J., Zank, L.A., Zank, G.P., 
1988, A\&A 198, 274

\bibitem[1989]{voelk_b} V\"olk, H.J., 1989, A\&A 218, 67

\bibitem[1986]{weiler} Weiler K.W., Sramek, R.A., 
Panagia, N., van der Hulst, J.M., Salvati,
1986, ApJ 301, 790


\bibitem[1994a]{xu_a} Xu, C., Lisenfeld, U., V\"olk, H.J., 
Wunderlich, E., 1994a, A\&A 282, 19

\bibitem[1994b]{xu_b} Xu, C., Lisenfeld, U., V\"olk, H.J., 1994b, A\&A 285, 19
\end{thebibliography}
\end{document}